\documentclass[aps,a4paper]{revtex4}

\usepackage[utf8]{inputenc}
\usepackage[english]{babel}
\usepackage{amsmath}
\usepackage{amsfonts}
\usepackage{amssymb}
\usepackage{graphicx}
\usepackage{wrapfig}  
\usepackage{hyperref}
\usepackage{mathrsfs}
\usepackage{float}
\usepackage{tikz}
\let\vec=\mathbf
\newcommand{\es}{{e_{\mathrm{s};\mathbf{kQ}}}}
\newcommand{\ea}{{e_{\mathrm{a};\mathbf{kQ}}}}
\newcommand{\easq}{{e^2_{\mathrm{a};\mathbf{kQ}}}}

\begin{document}

\title{Incommensurate magnetic order in rare earth and transition metal compounds with local moments}
\author{A.K.~Pankratova$^{1,2}$, P.A.~Igoshev$^{1,2}$, V.Yu.~Irkhin$^1$}

\affiliation{$^{1}$Institute of Metal Physics, 620108, Kovalevskaya str. 18, Ekaterinburg, Russia\\
$^{2}$Ural Federal University, 620000, Mira str. 132, Ekaterinburg, Russia
}
\date{\today}

\graphicspath{{fig/}}

\date{\today}

\begin{abstract}
Within the framework of the $s$-$d(f)$ exchange model in the mean-field approximation 
for square, simple cubic, body-centered and face-centered cubic lattices, the formation of a ferromagnetic, spiral, and commensurate antiferromagnetic order is investigated. 
The possibility of the formation of inhomogeneous states (magnetic phase separation), which necessarily arises during first-order phase  transitions in the electron filling parameter, is taken into account. The saturation of the antiferromagnetic and spiral states is studied depending on the parameters of the model. The results obtained include a rich variety of magnetic structures and phase transitions, allowing the interpretation of magnetic properties of semiconducting and metallic systems containing magnetic atoms.
\end{abstract}
\maketitle

\section{Introduction}
The study of magnetic order in $d$- and $f$-systems is an important area of quantum solid state physics.
Magnets with well-defined local moments include pure rare-earth metals \cite{Vonsovskii:Magnetism,II} and, with reservations, transition $d$ metals of the iron group~ \cite{Vonsovskii:Magnetism,2015:Igoshev.a-Fe}, as well as many of their compounds.
In recent years, magnetic systems with unusual properties, including compounds with intermediate valence and heavy fermions \cite{II,I17}, have been actively studied. Such systems are usually described within the framework of the many-electron models, the $s$-$d(f)$ exchange model and the periodic Anderson model as the Kondo lattices where Kondo screening~\cite{Kondo_effect} of magnetic moments by conduction electrons occurs.

Both experimental and theoretical studies show \cite{601,IK,II,I17} that magnetic ordering, often with a reduced magnetic moment, is widespread among these compounds, which brings them close to itinerant magnets.
Depending on the ratio of the parameters in Kondo systems, both states with a very small moment and the with small suppression (as in pure metals) can be realized.
In these systems, a variety of forms of magnetic ordering occurs due to frustrations of long-range exchange interactions (usually this is an indirect Ruderman-Kittel-Kasuya-Yosida (RKKY) interaction via conduction electrons). Thus, many compounds of rare earth elements (especially cerium) have a particularly complex magnetic structure due to the combined influence of exchange interactions, crystal field, and Kondo screening~\cite{I17}. 

In this paper, we consider the spatial structure of ferro- and antiferromagnetic ordering for metallic magnets with weak screening of the magnetic moment. 

This situation is typical, besides a number of rare-earth compounds,  for many compounds of transition metals, such as manganese and chromium, whose atoms in the free state have a large moment which is mainly preserved in the metallic state. 
In manganese phosphide MnP, there is a competition between the helical and ferromagnetic orders, and the phase transition between them occurs under pressure increase~\cite{1966:MnP,1974:Helimagnetism,2016:MnP}. 
A completely different situation occurs for CrAs$_{1-x}$ Sb$_x$, where there is a competition of antiferromagnetic (AFM) and incommensurate (spiral, order of the ``double helix'' type) magnetic order, and the phase transition between them occurs at $x\sim0.5$~\cite{1974:Helimagnetism,2014:CrAs1,2014:CrAs2,2015:CrAs}.  
The FeP 
compound exhibits a double helix magnetic order with a critical temperature $T_{\rm N} = 125$~K~\cite{1971:FeP}. 
In this compound, large magnetoresistance effect, which is linear in magnetic field and  stable up to very  fields, was found~\cite{2021:FeP_Campbell}. The fact that phonon calculation  spectra in this compound can be calculated within standard density functional technique allows to suggest that magnetism in this compound is to a large extent itinerant~\cite{2021:FeP_Sukhanov}, despite the fact that non-trivial topological effects (Dirac point etc.) were caused by double helix magnetic order provided that the magnetic field is directed along $c$ axis~\cite{2021:FeP_Campbell}.  
In the FeAs compound, an incommensurate (spiral or spin density wave type) magnetic order~\cite{2011:FeAs} with a wave vector directed along the $z$ axis of the tetragonal B31 structure was found.
An exhaustive review of the dependence of the magnetic order on the constituent elements is given in the review~\cite{2016:Review_MnP}.
The  spiral order found in hole-doped cubic manganites $R$MnO$_3$ (realized at $R=$~Tb and Dy) competes with A-type antiferromagnetism (wave vector $\mathbf{Q}=(0,0,\pi/c)$) for $R=$~La, Pr, Nd, and Sm and E-type for $R=$~Ho~\cite{2003:Kimura,2005:Kimura}  and ferroelectric compound for~$R=$~Y \cite{1963:RMnO3_Yakel,1964:RMnO3_Bokov,2007:YMnO3_Lorenz} ($\uparrow\uparrow\downarrow\downarrow$ order in MnO$_2$ planes).

In ytterbium compounds, the observed magnetic order can vary considerably. 
The YbPdSi system is a heavy-fermion ferromagnet with the Curie temperature $T_{\rm C} = 8$~K~\cite{YbPdSi}. 
Study of  Yb$T$Ge ($T = $~Rh, Cu, Ag)~\cite{YbTGe} compounds showed that YbRhGe is an antiferromagnet with $T_{\rm N}=7$~K, YbCuGe is a ferromagnet with the moment $0.7\mu_{\rm B}$ and $T_{\rm C}=8$~K,
YbAgGe \cite{YbAgGe} demonstrate low-temperature non-collinear magnetic ordering depending on  temperature in a complicated way. In the ground state, in a latter compound a commensurate AFM order occurs with the wave vector $\mathbf{Q}=(1/3,0,1/3)$, and an increase in temperature leads to a transition to a state of incommensurate order $\mathbf{Q}=(0,0,0.324)$ at $\sim 0.7$~K through a first-order phase transition \cite{2004:Fak,2006:Fak}) as well as to non-Fermi-liquid behavior~\cite{YbAgGe1}. 
In the antiferromagnetic system  Yb(Rh$_{1-x}$Co$_x$)$_2$Si$_2$ a transition to ferromagnetism was discovered with doping at $x=0.27$ below $T_{\rm C}=1.3$~K, the moment being $0.1 \mu_{\rm B}$~\cite{DopedYbRh2Si2}. 
Yb$_2$(Pd$_{1-x}$Ni$_x$)$_2$Sn \cite{Yb2(Pd1?xNix)_2Sn} and Yb$_2$Pd$_2$In$_{1-x}$Sn$_x$ \cite{Yb2Pd2In1-xSnx} systems exhibit frustrated magnetism, the unusual magnetic ordering arising upon doping. In the latter system, magnetic ordering appears in a narrow region near $x=0.6$ on the background of non-Fermi-liquid behavior.

Several cerium compounds like Ce$TX$ ($T$ is a transition metal, $X$ is an element of $3-5$ main group of the periodic table) exhibit ferromagnetic ordering with a low Curie temperature~\cite{2016:Janka:CeTX}. 
Thus, the CePtAl compound demonstrates a number of magnetic phase transitions between phases with wave vectors $\mathbf{Q}=0$, $(0,0.463(8),0)$ and $(0,0.5,0)$~\cite{1995:CePtAl}. 
CeAgAl~\cite{2008:CeAgAl} and CeAgGa~\cite{2011:CeAgGa} compounds are ``Kondo'' ferromagnets, which is revealed by magnetic measurements (negative paramagnetic Weiss temperature in the Curie-Weiss law). 
Neutron diffraction experiments in the Kondo compound CePdSn confirm spiral magnetic order with a wave vector $\mathbf{Q} = (0, 0.473, 0)$~\cite{1992:CePdSn.Kasaya}. 
Apparently, the applied pressure has a significant effect on this magnetic order, so that at pressures above 6 GPa a~sharp feature of the electrical resistance disappears~\cite{1993:CePdSn_YbB12.Iga}.  

CePtSn is ordered antiferromagnetically (there are two close phase transitions with critical temperatures of 5.5 and~7.8~K).
The effective magnetic moment $\mu_{\rm eff}=2.85 \mu_{\rm B}$ turns out to be larger than expected, and the Weiss constant
$\theta_{\rm p}=-9.2$~K is in a good agreement with the critical temperature~\cite{1995:CePtSn.TbPtSn.Kolenda}.  
Neutron scattering experiments at $ T = 1.7$~K show additional reflections confirming antiferromagnetism~\cite{1994:CePtSn.Adroja}. 
The magnetic order parameters are determined at $T=1.7$~K. The wave vector is $ \mathbf{Q} = (0, 0.428, 0) $, and the magnetization amplitude is directed along the axis $c$.
The wave vector changes with temperature. Above the temperature $T<3.8$~K there is an incommensurate antiferromagnetic order with $\mathbf{Q} = (0, 0.418, 0)$~\cite{1993:CePtSn.Kadowaki}.

It is usually assumed that the magnetic order in $d$- and $f$-compounds  is formed as a result of the Heisenberg exchange interaction between localized electron spins. Within the framework of this approach, the explanation of the complex non-collinear or spiral magnetic order in these compounds requires  the existence of anisotropic relativistic interactions \cite{Coqblin,1988}.  
An alternative approach is to consider the $s$-$d(f)$ exchange as the cause of the spiral magnetic order in the absence of relativistic interactions.
The observed variety of properties of transition metal compounds is determined by the properties of the subsystem of itinerant electrons, i.e. by their spectrum determined by the geometry of bonds  in the lattice, as well as by the value of the parameter of the $s$-$d(f)$-exchange coupling $I$ between the subsystem of localized and itinerant electrons. Physically, the interaction (and hence the magnetic order) between local moments is realized  through the transfer in the subsystem of itinerant electrons. Concrete form of the magnetic order is determined, besides the interaction strength $I$, by the form of the Fermi surface. 

Another important consequence of the presence  of itinerant-electron subsystem is the appearance of inhomogeneous states arising at any first-order phase transition in metallic systems. In recent years, phase separation in cuprates~\cite{1994:Dagotto,1995:Tranquada,1995:Nagaev,2001:Nagaev}, and in manganites~\cite{2001:Kagan} has attracted a lot of attention, including theorical approaches~\cite{2010:Igoshev,2015:Igoshev}. However, for the problem of the  magnetic order formation in the $s$-$d(f)$ model this aspect has not received sufficient attention. 

In this paper, we study the phase diagrams of the magnetic state depending on the $s$-$d(f)$ exchange parameter. 
We will also perform comparison with the previously obtained results in the framework of the one-band Hubbard model~\cite{2010:Igoshev,2015:Igoshev,2017:Igoshev}, as well as in  the Anderson model~\cite{2016:Igoshev} where the effects of non-integer filling $d$-orbitals can play an essential role.

\section{Method}

The Hamiltonian of the $s$-$d(f)$ model is
\begin{equation}\label{eq:Hamilt_start}
\mathscr{H} = \sum \limits _{ij \sigma} t _{ij}  c_{i \sigma} ^{\dagger} c_{j \sigma} - \dfrac{I}{2} \sum \limits _{i \sigma \sigma '} \left( \vec{S}_i \overrightarrow{\sigma}_{\sigma\sigma '} \right) c_{i \sigma} ^{\dagger} c_{i \sigma'}
\end{equation}
where $c_{i \sigma} ^{\dagger} (c_{i \sigma})$ is the creation (annihilation) operator of an itinerant $s$-electron at site $i$ with spin projection $\sigma = \uparrow, \downarrow$, $\overrightarrow{\sigma}$ is the vector of Pauli matrices, $\mathbf{S}_i$ is the spin operator of localized $d(f)$ electrons, 
\begin{gather}
t_{ij} = \dfrac{1}{N} \sum \limits _{\vec{k}} \varepsilon (\vec{k}) e ^{\mathrm{i}\vec{k} \left( \vec{R}_i - \vec{R}_j \right)},
\end{gather}
are the matrix elements of the electron transfer, where $\varepsilon(\vec{k})$ is the spectrum of $s$-electrons in the absence of interaction. We choose the spectrum in the tight-binding approximation, taking into account the nearest (integral $t_{ij}=-t$) and next-nearest (integral $t_{ij}=+t'$) neighbor hopping integrals. 

We assume that the  magnetization of $s$- and $d$-electron states form spiral magnetic order in the $yz$-plane 
\begin{eqnarray}
    \mathbf{m}^s_{i} &=& m_s (\sin (\mathbf{QR}_i)\mathbf{e}_y + \cos (\mathbf{QR}_i)\mathbf{e}_z), \\
    \mathbf{m}^d_{i} &=& m_d (\sin (\mathbf{QR}_i)\mathbf{e}_y + \cos (\mathbf{QR}_i)\mathbf{e}_z),
\end{eqnarray}
where $m_s$ and $m_d$ are amplitudes of the magnetization, $\mathbf{e}_{y(z)}$ is a unit vector of $y(z)$ axes. 

When passing to a local coordinate system where the $z$-axis is aligned in the direction of magnetization at each site (local rotation through an angle $\mathbf{QR}_i/2$ around the $x$-axis), the operators $c_{i\sigma}$ are transformed as
\begin{equation}
    c_{i\sigma} \rightarrow \sum_{\sigma'} (\cos (\mathbf{QR}_i/2)\delta_{\sigma\sigma'} + \sin (\mathbf{QR}_i/2)\sigma^x_{\sigma\sigma'}) c_{i\sigma'}.
\end{equation}
(The operator $\mathbf{S}_i$ transforms correspondingly.) 
In the new coordinate system, the magnetization amplitude for both subsystems is directed along the $z$-axis at the cost of the  transfer being non-diagonal in the spin index. The amplitudes of the magnetization of $s$- and $d(f)$-electrons are determined by the relations
\begin{gather}
m_s = \dfrac{1}{2} \sum \limits _{\sigma} \sigma \langle c_{i \sigma} ^{\dagger} c_{i \sigma} \rangle, \\
m_d = \langle S^z_i \rangle = 1/2,
\end{gather}
since we neglect local momentum fluctuations. The angle brackets
\begin{equation}
\langle\cdots\rangle = \frac1{Z}{\rm Tr}\exp(-(\mathscr{H} - \mu\mathscr{N}_s)/T) 
\end{equation}
correspond to the standard statistical averaging within the grand canonical ensemble,  $\mu$~is the~chemical potential, $\mathscr{N}_s$~is the operator of the number of $s$-electrons.

Below we use the mean-field approximation, which corresponds to replacing the spin operator $\mathbf{S}_i$ by a vector with a given configuration (in our case, this is $\mathbf{e}_z/2$), thereby neglecting the fluctuations of localized electron spin $\delta\mathbf{S}_i = \mathbf{S}_i - \mathbf{e}_z/2$ and is applicable for the case of zero temperature and weak Kondo screening of the local moments.

Applying the transformation to the Bloch states, $c_{i\sigma} = N^{-1/2}\sum_{\mathbf{k}}c_{\mathbf{k}\sigma}\exp(-\mathrm{i}\mathbf{kR}_i)$, we get the Hamiltonian  
\begin{equation} 
\mathscr{H} _{\rm MF} = \sum \limits _{\vec{k} \sigma} \left( \left( \es - \dfrac{I}{4} \sigma \right) c_{\vec{k} \sigma} ^{\dagger} c_{\vec{k} \sigma} + \ea  c_{\vec{k} \overline{\sigma}} ^{\dagger} c_{\vec{k} \sigma} \right),
\end{equation}
where $e_{{\rm s,a};\vec{kQ}} = \dfrac{1}{2} \left( \varepsilon(\vec{k} + \vec{Q}/2) \pm \varepsilon(\vec{k} - \vec{Q}/2) \right).$

As a result of diagonalization 
\begin{eqnarray}
    c_{\vec{k}\uparrow} &=& \cos\theta_{\vec{k}}a_{\vec{k}1} - \sin\theta_{\vec{k}}a_{\vec{k}2}, \\
    c_{\vec{k}\downarrow} &=& \sin\theta_{\vec{k}}a_{\vec{k}1} + \cos\theta_{\vec{k}}a_{\vec{k}2}, 
\end{eqnarray}
with $\theta_{\vec{k}} = -\frac12\arctan ({4\ea}/I)$, we obtain  the Hamiltonian 
\begin{equation}
    \mathscr{H}_{\rm MF} = \sum_{\mathbf{k}\nu} E_{\nu\vec{Q}}(\vec{k})a^\dag_{\mathbf{k}\nu}a_{\mathbf{k}\nu},
\end{equation}
where the branches of AFM spectrum are labelled by $\nu = 1, 2$ and read 
\begin{equation}\label{eq:Ek}
E_{\nu\vec{Q}}(\vec{k}) = \es + (-1)^\nu\sqrt{\left(I/4 \right) ^2 + \easq}.
\end{equation}

The concentration and $s$-electrons magnetization amplitude are determined by the equations
\begin{eqnarray}
\label{eq:ns}
n_s \left( \vec{Q}, \mu \right) &=& \dfrac{1}{N} \sum \limits _{\vec{k}\nu} f \left( E_{\nu\vec{Q}}(\vec{k}) \right), \\
\label{eq:ms}
m_s \left( \vec{Q}, \mu \right) &=& \dfrac{1}{2N} \sum \limits _{\vec{k}\nu} (-1)^{\nu+1} \cos 2\theta_{\vec{k}}f \left( E_{\nu\vec{Q}}(\vec{k})\right),
\end{eqnarray}
where $f(E) = (\exp((E - \mu)/T) + 1)^{-1}$ is the Fermi function. 

When constructing the magnetic phase diagram of the ground state~($T=0$)
for a given value of $\mu$ from the equation (\ref{eq:ns}), (\ref{eq:ms}), we determined $n_s$ and $m_s$ for all possible spiral phases, while $\mathbf{Q}$ was changed on a sufficiently (adaptively selected) dense grid, see~\cite{2010:Igoshev,2015:Igoshev} for details of the method. Numerical  quasimomentum integration was carried out within the framework of the improved tetrahedron method~\cite{1994:Andersen}.
In this case, a phase is realized that corresponds to the minimum value of the grand potential $\Omega \left( \vec{Q}_{\rm min}, \mu, m_{\rm min}, n_{\rm min} \right)$ 
\begin{equation} \label{eq:omega}
\Omega_{\mathbf{Q}}(\mu) = \frac1{N}\sum \limits _{\vec{k} \nu} \left( E_{\nu\vec{Q}}(\vec{k}) - \mu \right) f \left( E_{\nu\vec{Q}}(\vec{k}) \right).
\end{equation}
This procedure being repeated on a fairly dense grid in $\mu$ allows one to construct a magnetic phase diagram of the ground state taking into account inhomogeneous states (phase separation) arising when the phase parameters $(\vec{Q}_{\rm min},m_{\rm min}, n_{\rm min}) $ experiences a jump as a function of $\mu$.

\section{Results}
Now we consider the resulting phase diagrams of the ground state in the variables $n_s$~---~$I$ for square, simple, body- and face-centered cubic lattices in the nearest and next-nearest neighbors approximation. For ``bipartite'' lattices  (which can be split into two sublattices with all nearest neighbors of one sublattice belonging to other), there is a particle-hole symmetry. Thus there is the transformation $c^{}_{i \sigma} \rightarrow (-1)^ic^\dag_{i \sigma} $ which corresponds to replacing particles by holes: $n_s\rightarrow (2 - n_s) $ and $ t '\rightarrow (-t') $, while the magnetic order is preserved. Then one can consider only the case $t'>0$, and in the case $t'=0$, consider the region $0<n_s\le1$.

\begin{wrapfigure}[20]{l}{0.3\textwidth}
\noindent
\begin{tikzpicture}[>=latex, scale=1.0]
\draw (-2,-2) -- (2,-2) -- (2,2) -- (-2,2) -- cycle;
\draw[->] (-2.5,0) -- (2.5,0);
\draw[->] (0,-2.5) -- (0,2.5);
\draw[thick,brown!80!red] (0,0) -- (2,0);
\draw[thick,brown!80!red] (0,0) -- (2,2);
\draw[thick,brown!80!red] (2,0) -- (2,2);
\draw[fill=red] (0,0) circle (2pt);
\draw[fill=red] (2,0) circle (2pt);
\draw[fill=red] (2,2) circle (2pt);
\node[below left] at (0,0) {\color{red}$\Gamma$};
\node[below right] at (2,0) {\color{red} X};
\node[above right] at (2,2) {\color{red} M};
\node[left] at (0,2.5) {k$_{\text{y}}$};
\node[above] at (2.5,0) {k$_{\text{x}}$};
\node[above] at (1,0) {\color{red}$\Delta$};
\node[right] at (2,1) {\color{red}$Z$};
\node[above left] at (1,1) {\color{red}$\Sigma$};
\end{tikzpicture}
\caption{The first Brillouin zone of a square lattice showing highly symmetric points and directions: $\Gamma(0,0)$, X$(\pi,0)$, M$(\pi,\pi)$, $\varSigma(q_x,q_x)$, $Z(\pi,\pi - \delta q_y)$.  
}
\label{fig:sqZB}
\end{wrapfigure}
By analogy with the saturation of the ferromagnetic state, we can introduce the concept of saturation for spiral and antiferromagnetic phases:
if the Fermi energy is in only one AFM subband, we will call such a saturated state, and otherwise unsaturated.

The property of saturation of AFM ordering, introduced in this way, is important for problems of electron and spin transport (in particular, in the saturated case, there is no contribution from one-magnon scattering processes). 

For all the phase diagrams found, a small number of carriers (physically, this case corresponds to the case of a degenerate semiconductor) favors the formation of ferromagnetic order, and an increase of $I$ leads to an expansion of the region of the ferromagnetic phase. The nonzero transfer integral $ t '$ between the second neighbors for bipartite lattices leads to an expansion (narrowing) of the ferromagnetic order region for $n_s \ll 1$ ($2 - n_s \ll 1$). 

For small $I$, the type of magnetic order is determined by the magnitude of the response to an infinitesimal modulated magnetic field; a magnetic order is formed corresponding to the wave vector that provides the maximum of the Lindhardt function
\begin{equation}
    \chi(\mathbf{Q}) = \frac1{N}\sum_{\mathbf{k}}\frac{f(\varepsilon(\mathbf{k} - \mathbf{Q}/2)) - f(\varepsilon(\mathbf{k} + \mathbf{Q}/2))}{\varepsilon(\mathbf{k} + \mathbf{Q}/2) - \varepsilon(\mathbf{k} - \mathbf{Q}/2)},
\end{equation}
which determines the indirect RKKY interaction of localized moments. 

\subsection{Square lattice}

\begin{figure} [ht]
\includegraphics[angle=-90,width=0.33\textwidth]{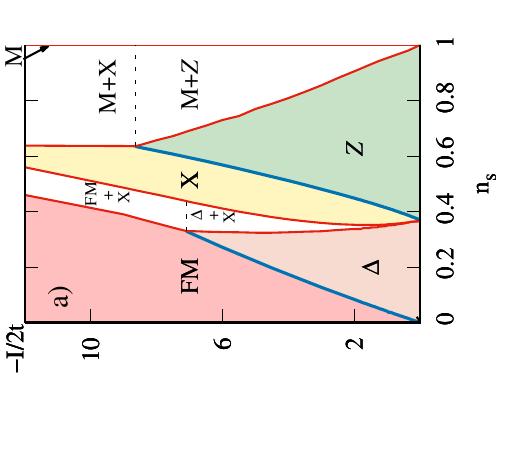}
\includegraphics[angle=-90,width=0.66\textwidth]{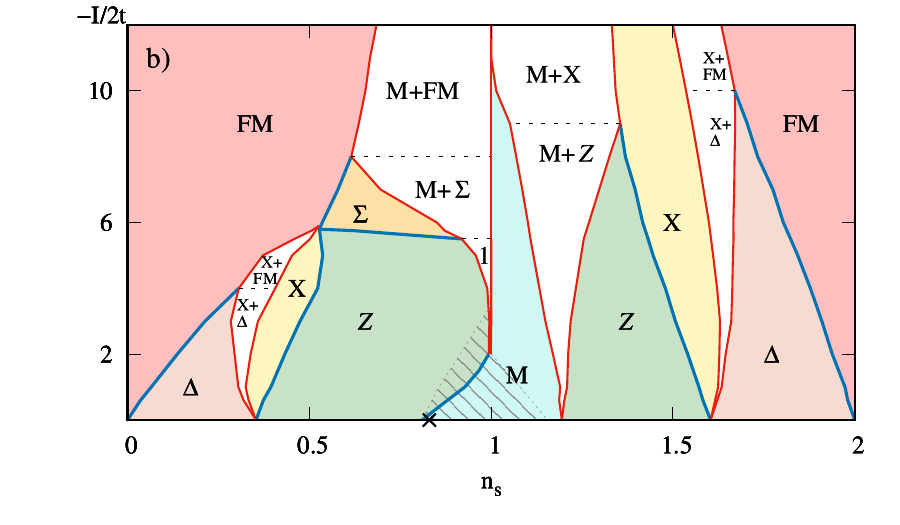}
\caption{Magnetic phase diagram of the ground state for a square lattice in the variables $ n_s $ --- $ I $, a) $ t '= 0 $, b)~$ t' = 0.2t $. The thick (blue) line indicates the lines of second-order phase transitions, the thin (red) line indicates the first-order phase transitions. White areas are phase separation areas, shaded areas are areas of unsaturated states. Label 1 denotes small \textit{Z}+M PS region. The cross at filling axis labels the value $n^{\rm sq}_{\rm vH} \approx 0.83$ corresponding to the coincidence of the van Hove singularity and the Fermi level. }
\label{fig:sq}
\end{figure}
Fig. \ref{fig:sq} shows the magnetic phase diagrams of the ground state for a square lattice at (a) $t'=0$ and~(b)~$t'=0.2t$, see~notation for highly symmetric points and directions in~Fig.~\ref{fig:sqZB}. The ferromagnetic order with a small number of carriers is saturated (half-metallic ferromagnetism \cite{IK94}) and is unstable with respect to an incommensurate spiral order with a wave vector in the $\Delta $-direction. For $ n_s = 1 $, the Néel antiferromagnetic order~(M, ``checkerboard'' order) dominates at an arbitrarily small value of~$ I $. In the case $ t'= 0 $, doping for small $ I \lesssim 8t $ leads to a first-order transition (through phase separation) to the phase with a wave vector  $ \mathbf{Q}_Z = (\pi, \pi - \delta q_y )$, and further doping leads to gradual motion of the wave vector of magnetic order along the line $ Z $ to $ \mathbf{Q} = \mathbf{Q}_{\rm X} = (\pi, 0) $. At $ 8t \lesssim I $, the first-order transition immediately occurs to the X-phase (through the phase separation). Further doping leads to a jump of the wave vector from $ \mathbf{Q}_ {\rm X} $ to the vector lying in the $ \Delta $-direction (wave vector $ \mathbf{Q} = (q_x, 0,0)$). The magnitude of  the wave vector jump  grows rapidly with increasing $ I $, so that at $I\gtrsim 7t$ the jump occurs immediately into the ferromagnetic phase.

\begin{figure}[ht]
\noindent
\includegraphics[angle=-90,width=0.49\textwidth]{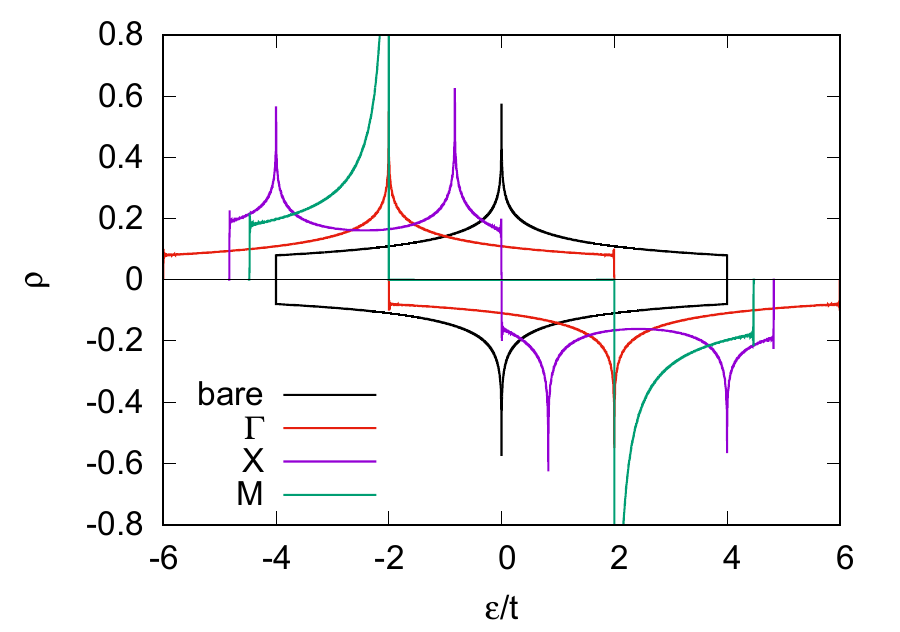}
\caption{The density of states (\ref{eq:DOS_def}) for spectrum $\epsilon(\mathbf{k})$ (labelled as ``bare'') and different $\mathbf{Q}$ ($\mathbf{Q} = 0,\mathbf{Q}_{\rm X},\mathbf{Q}_{\rm M}$) subbands of collinear AFM spectrum of $s$-states $E_{\nu\vec{Q}}(\vec{k})$, see Eq.~(\ref{eq:Ek}), in units of $t^{-1}$ for the square lattice at $t' = 0$ and $I = 4t$. The density of states of lower (upper) AFM subband is shown in upper (lower) half-plane.}
\label{fig:DOS_sq}
\end{figure}
The inclusion of a finite $t'$ (due to symmetry, for bipartite lattices we consider only the case $ t'> 0 $, see~above) leads to an asymmetry of the phase diagram owing to asymmetry of the bare spectrum and the influence of the strong van Hove singularity which is not smeared  at finite $t'$. In particular, for $ n_s<1 $ the region of ferromagnetism increases and a new  (in comparison with the case $ t'=0 $)  phase is realized with a wave vector lying on the diagonal of the Brillouin zone ($ \varSigma $-direction).
In the case $ n_s>1 $, the Néel antiferromagnetic order is already realized (instead of phase separation) near the half-filling. The \textit {unsaturated} antiferromagnetic order is realized only for small $ I $ near the half-filling.  
The phase separation region is extensive near the half-filling and increases as $ I $ grows. All the states, except for a small area on the diagram in the next-nearest neighbor approximation, are saturated.

The density of states (DOS) at $t' = 0$ for the bare spectrum $\varepsilon(\mathbf{k})$, as well as  for the spectrum  $E_{\nu\vec{Q}}(\vec{k})$   
\begin{equation}\label{eq:DOS_def}
\left(
\begin{array}{c}
\rho(\varepsilon)\\
\rho_{\mathbf{Q},\nu}(\varepsilon)
\end{array}
\right)=
\frac1{N}\sum_{\mathbf{k}}
\left(
\begin{array}{c}
\delta(\varepsilon - \varepsilon(\mathbf{k}))\\
\delta(\varepsilon - E_{\nu\mathbf{Q}}(\mathbf{k}))
\end{array}
\right)
\end{equation}
for different $\mathbf{Q}=\mathbf{Q}_{\Gamma}=0 (\text{corresponding to ferromagnetism}),\mathbf{Q}_{\rm X},\mathbf{Q}_{\rm M}$
was calculated using the tetrahedron method\cite{1994:Andersen} and is shown in Fig.~\ref{fig:DOS_sq}. It is clear that the van Hove singularity is retained after splitting of the spectrum in the ordered state. The gap and the most strong van Hove singularity is present for the case of perfect nesting property,
\begin{equation}\label{eq:nesting_def}
\varepsilon(\mathbf{k} + \mathbf{Q})	= -\varepsilon(\mathbf{k})
\end{equation}
at $\mathbf{Q} = \mathbf{Q}_{\rm M}$.

Strictly speaking, filling corresponding to the van Hove singularity at $t' \ne 0$ does not coincide with $n_s = 1$ (an~example of the position corresponding to van Hove filling is shown by the cross in Fig.~\ref{fig:sq}b), which stabilizes the ferromagnetic ordering at infinitely small $I$. The presence of ferromagnetic ordering at finite $I$ is a difficult issue, since for Fermi level being in the vicinity of the logarithmic van Hove singularity, there is a delicate competition of ferromagnetic ordering  and surrounding AFM and spiral ordering. 
We can state that ferromagnetic ordering generally occupies a small region of the phase diagram, even in the mean-field approximation, and is probably unstable with respect to correlation effects, see arguments for the similar problem within the Hubbard model in~\cite{2003:Kampf,2011:fRG_Igoshev}.

\subsection{Simple cubic lattice}
\begin{wrapfigure}[21]{l}{0.3\textwidth}
\noindent
\includegraphics[angle=-90,width=0.29\textwidth]{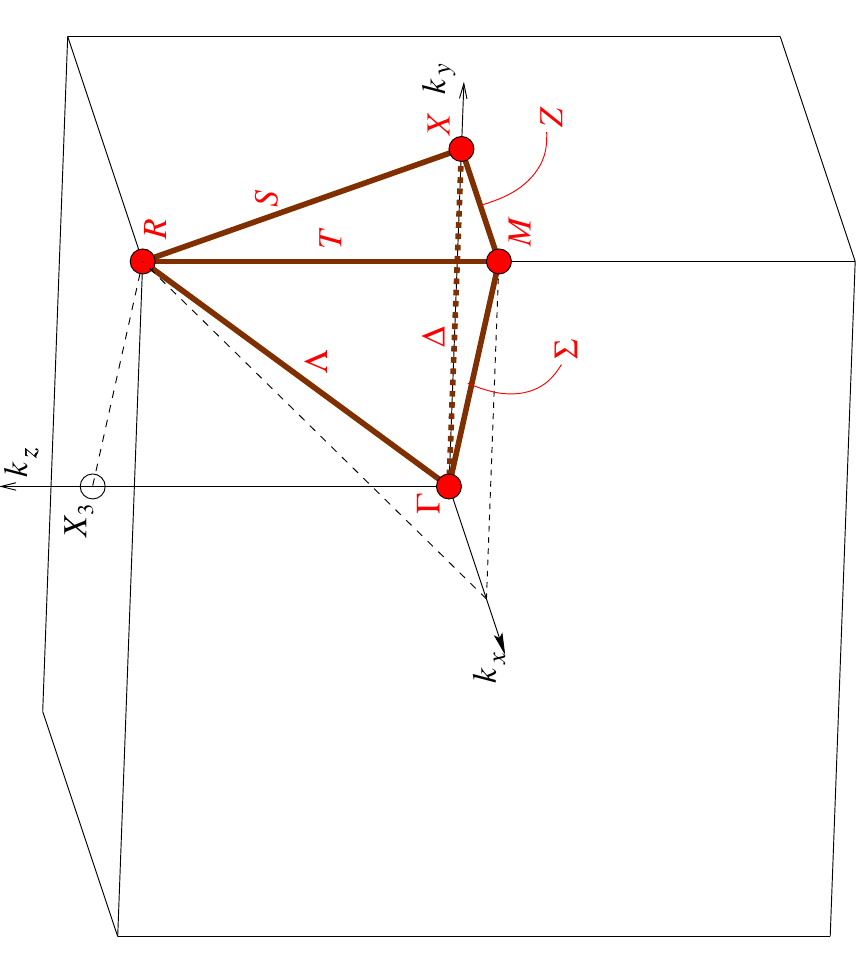}
\caption{The first Brillouin zone of a simple cubic lattice showing highly symmetric points and directions: $\Gamma(0,0,0)$, X$(\pi,0,0)$, M$(\pi,\pi,0)$, R$(\pi,\pi,\pi)$, $\varDelta(q_x, 0, 0)$, $\varSigma(q_x, q_x, 0)$, $\varLambda(q_x, q_x, q_x)$, $Z(\pi,\pi - \delta q_y, 0)$, $T(\pi,\pi, \pi - \delta q_z)$, $T(\pi,\pi - \delta q_y, \pi - \delta q_y)$.}
\label{fig:scZB}
\end{wrapfigure}
Fig.~\ref{fig:sc} shows the magnetic phase diagram of a simple cubic lattice with (a) $ t'= 0 $, (b) $ t'=0.2t $; , see~notation for highly symmetric points and directions in~Fig.~\ref{fig:scZB}.
As might be expected, the phase diagram at $ t'=0 $ is similar to that for a square lattice (Fig.~\ref{fig:sq}a).
A significant difference 
is that the region of the ferromagnetic phase, even at extremely small $ I $, extends up to $ n_s \sim 0.1 $,  
although the ferromagnetic order immediately becomes unsaturated upon doping. 

\begin{figure}[b]
\includegraphics[angle=-90,width=0.49\textwidth]{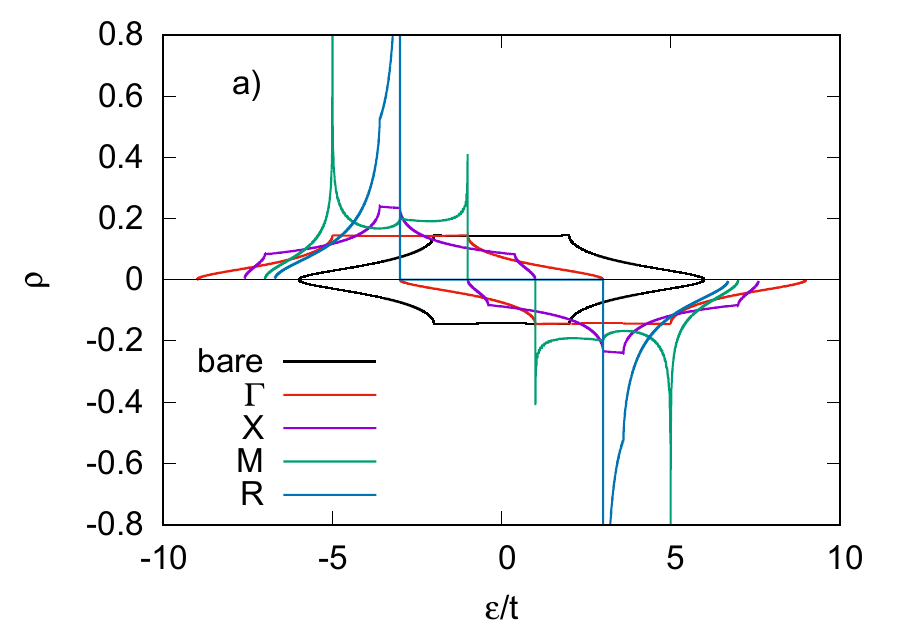}
\includegraphics[angle=-90,width=0.49\textwidth]{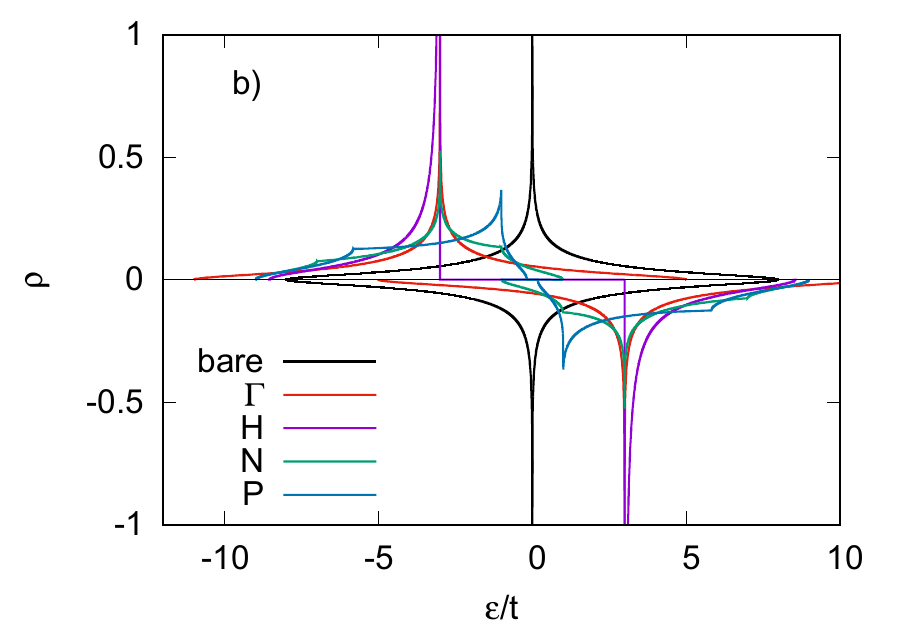}
\caption{The same as in Fig.~\ref{fig:DOS_sq} at $I = 6t$ and $t' = 0$ for a) sc lattice ($\mathbf{Q} = 0,\mathbf{Q}_{\rm X},\mathbf{Q}_{\rm M}, \mathbf{Q}_{\rm R}$), b) bcc lattice ($\mathbf{Q} = 0,\mathbf{Q}_{\rm H},\mathbf{Q}_{\rm N}, \mathbf{Q}_{\rm P}$).}
\label{fig:DOS_sc_and_bcc}
\end{figure}
An increase of doping at small $ I \lesssim 7t $ leads to a jump transition from the ferromagnetic state to the X antiferromagnetic phase (ferromagnetically ordered planes which are  antiferromagnetically ordered with respect to each other, collinear AFM order of type~A in terms of~\cite{book:Goodenough}). Further doping leads to a change in the wave vector through a first-order transition (and phase separation) from X to M (for small $ I \lesssim 5t $, through the unsaturated $ Z $ phase). In terms of \cite{book:Goodenough}, the M order is a collinear AFM order of type~C. For  these types of magnetic orders, whole planes of the lattice are ordered in the same way.
In the vicinity of  half-filling $ n_s = 1 $, a phase separation occurs into the Néel antiferromagnetic (checkerboard) order~(R, $ \mathbf{Q} _ {\rm R} = (\pi, \pi, \pi) $,  collinear AFM order of type~G in terms of~\cite{book:Goodenough}), which is caused by the nesting of the spectrum, see Eq.~(\ref{eq:nesting_def}), and the $ T $-spiral incommensurate order, $ \mathbf{Q}_ {T} = (\pi, \pi, \pi - \delta q_z) $. In this respect, the phase diagram is similar to the corresponding diagram for a square lattice (the R-order being similar to the M-order for a square lattice, and the $ T $-order to the $ Z $-order). When $ n_s $ is not too large ($ \lesssim 0.5 $), the antiferromagnetic (including spiral) order is always unsaturated.

Enabling transfer between second neighbors (Fig.~\ref{fig:sc}b) leads to the following asymmetric changes in the phase diagram. At $ n_s<1$ half of the phase diagram, a spiral phase is formed with a wave vector directed along the diagonal of the Brillouin zone~($\varLambda$-phase). This phase largely displaces the $T$-phase at intermediate values of $ I \lesssim 8t $ and intermediate doping $ n_s \gtrsim 0.5 $. In a narrow vicinity of the half-filling, a~pure R-phase is formed, which is replaced by the $ T $ phase outside this region. In a narrow vicinity of $ n_s = 2 $, a region  of the antiferromagnetic $ \Delta $-phase appears. The saturation of the ferromagnetic phase in the semiconductor limit essentially depends on the sign of $ t'$: for $ n_s<1 $ the order will be saturated starting from small $ I $ in a wide region of the phase diagram, while for $ n_s>1 $ the ferromagnetic order is unsaturated up to large $ I \sim 6t $. 

\begin{figure}[b]
\includegraphics[angle=-90,width=0.33\textwidth]{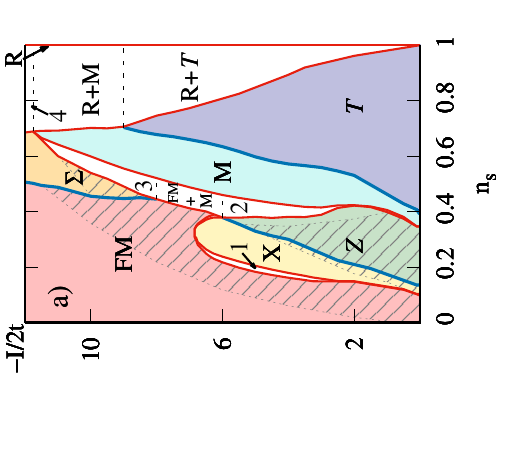}
\includegraphics[angle=-90,width=0.66\textwidth]{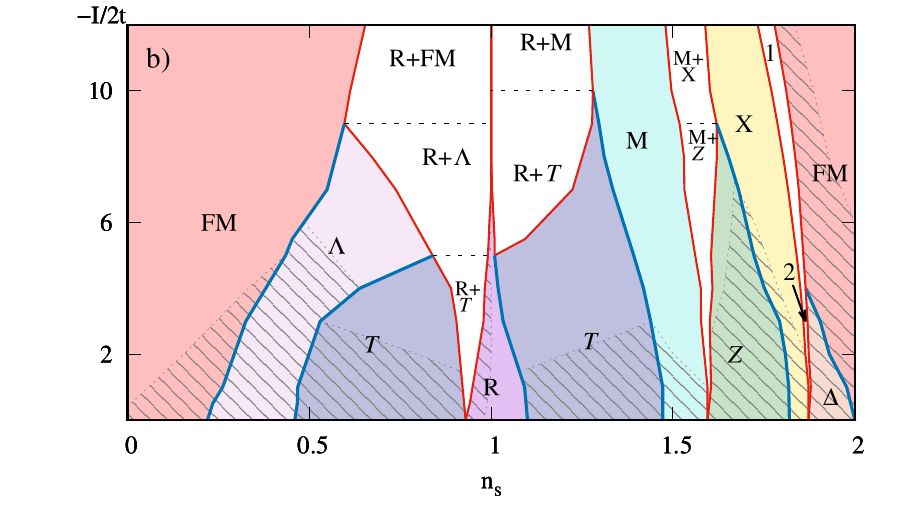}
\caption{The same as in Fig.~\ref{fig:sq} for a simple cubic lattice. a) $t'=0$ with labels corresponding PS regions: 1 --- FM+X, 2 --- \textit{Z}+M, 3  --- $\varSigma$+M, 4 --- R+$\varSigma$, b) $t'=0.2t$ with labels corresponding PS regions:  1 --- FM+X, 2 --- X+$\varDelta$.}
\label{fig:sc}
\end{figure}

The plots of  DOS for AFM subbands are shown in Fig.~\ref{fig:DOS_sc_and_bcc}a.
One can see a strong square root van Hove singularity for the $\mathbf{Q}=\mathbf{Q}_{\rm R}$ point owing to nesting,  and a more weak van Hove line singularity with diverging mass~\cite{2019:Igoshev_DOS_JETP,2019:Igoshev_DOS_FMM} for the $\mathbf{Q}=\mathbf{Q}_{\rm M}$ point.

\subsection{BCC lattice}
\begin{wrapfigure}[22]{l}[0pt]{0.3\textwidth}
\noindent\includegraphics[angle=-90,width=0.29\textwidth]{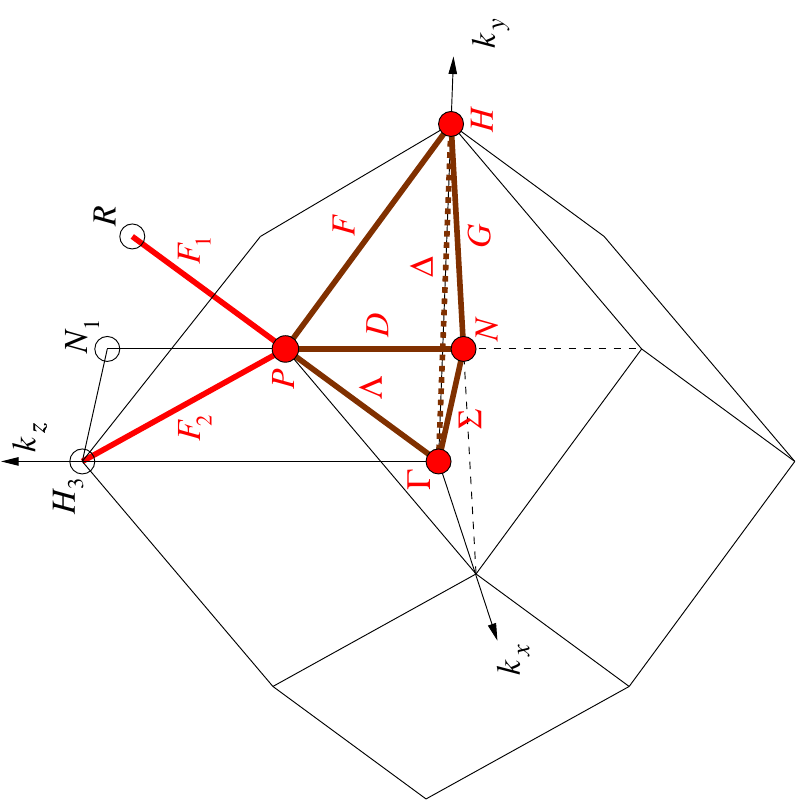}
\caption{The first Brillouin zone of a bcc lattice showing highly symmetric points and directions: $\Gamma(0,0,0)$, H$(2\pi,0,0)$, N$(\pi,\pi,0)$, P$(\pi,\pi,\pi)$, $\varDelta(q_x, 0, 0)$, $\varSigma(q_x, q_x, 0)$, $\varLambda(q_x, q_x, q_x)$, $D(\pi,\pi - \delta q_y, 0)$, $G(2\pi - \delta q_x, \delta q_x, 0)$, $F(2\pi - \delta q_x, \delta q_x, \delta q_x)$.}
\label{fig:bccZB}
\end{wrapfigure}
\begin{figure} [b]
\includegraphics[angle=-90,width=0.33\textwidth]{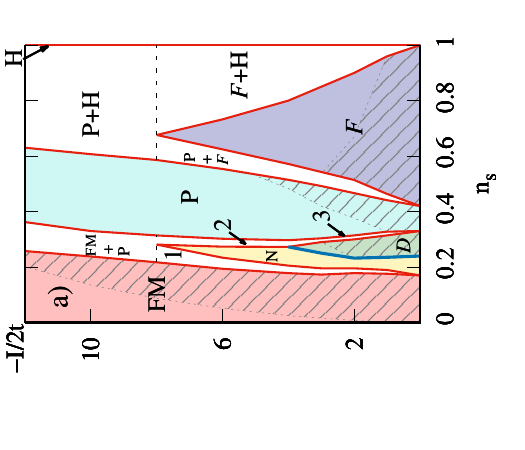}
\includegraphics[angle=-90,width=0.66\textwidth]{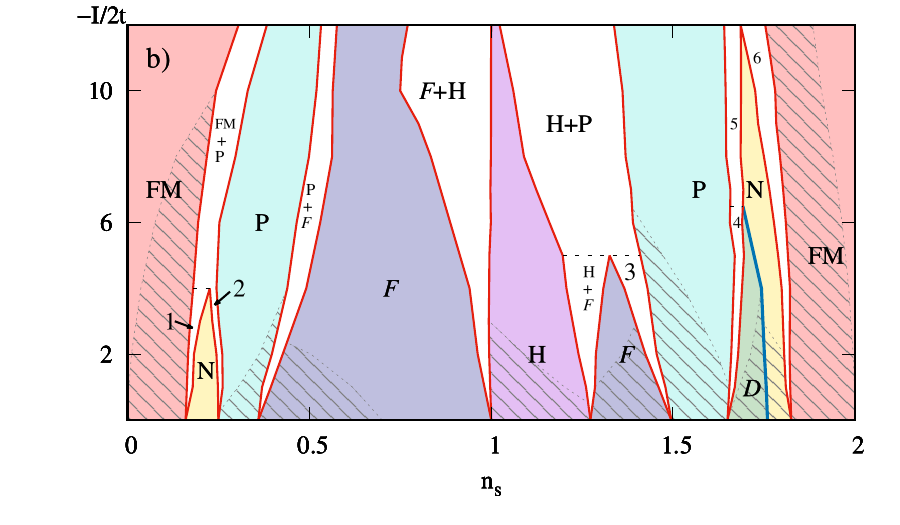}
\caption{The same as in Fig. \ref{fig:sq} for bcc lattice. a) $t'=0$ with labels corresponding PS regions: 1  --- FM+N, 2 --- N+P, 3 --- \textit{D}+P, b) $t'=0.2t$ with labels corresponding PS regions: 1  --- FM+N, 2 --- N+P, 3 --- \textit{F}+P, 4 --- P+\textit{D}, 5 --- P+N, 6 --- N+FM
.}
\label{fig:bcc}
\end{figure}

The magnetic phase diagram of the ground state for a bcc lattice is shown in Fig.~\ref{fig:bcc}: (a) $ t'= 0 $, (b) $ t'= 0.2t$, see~notation for highly symmetric points and directions in~Fig.~\ref{fig:bccZB}. Main types of magnetic order are: ferromagnetic with a small number of carriers, diagonal commensurate spiral magnetic order (P-phase, $ \mathbf{Q} _{\rm P} = (\pi, \pi, \pi) $), a similar order with a wave vector on the diagonal of the Brillouin zone edge $ D $ (including wave vector N corresponding to collinear AFM of the third type according to~\cite{book:Goodenough}) at intermediate $ n_s \sim 0.4 $, and strong Néel (checkerboard) order owing to the $ \log^2 {t}/{\varepsilon}$ van Hove singularity at the center of the zone (H phase, $ \mathbf{Q} _{\rm H} = (2 \pi, 0, 0) $, collinear antiferromagnetism of the first type according to~\cite{book:Goodenough}). The~stability of~the~latter order is responsible for wide regions of phase separation with the participation of the H-phase. 
Thus, we can conclude that the wave vector of the antiferromagnetic phase usually lies on the edge of the Brillouin zone, except for the case of a small number of carriers (``semiconductor limit''). The formed magnetic order is saturated, except for the cases of ferromagnetic ordering and of a small interaction $ I $, which is due to the tendency towards antiferromagnetism in the vicinity of half-filling.
Collinear second-type antiferromagnetism according to~\cite{book:Goodenough} is not taken into account in our calculations. This is essentially connected with the introduction of four sublattices and corresponds to the wave vector $ \mathbf{Q} _{\rm P} $; by analogy with Ref.~\onlinecite{2016:Timirgazin_MIT_for_FCC}, one can expect that it is realized at large $t'$ only.

For $ t'=0 $, the deviation of $ n_s $ from half-filling leads to the formation of an order of the $ F $-type ($ F $ is a line connecting points P and H) and a phase separation between H and $ F $ for not too large $ I $. For $ n_s \sim 0.4 $ and small $ I $, N- and $ D $-magnetic orders are formed. An increase of $ I $  leads to occurrence of three types of commensurate order: ferromagnetism for small $ n_s $, P for intermediate $ n_s $, and phase separation into P- and H-phases in a  wide vicinity of half-filling.

The nonzero transfer integral $ t'$ determines the asymmetry of the phase diagram. For $ n_s <1 $, as in the cases considered above, it leads to the appearance of a significant region of the $ F $-phase extending up to large $ I $, which participates in phase separation in the vicinity of $ n_s = 1 $~(the width of the separation region is considerably more narrow). In this case, the N-phase region decreases in height, remaining rather narrow. In the case of $ n_s> 1 $, a pure H-phase is formed, which, with an increase in $ n_s $, participates in the  phase separation with the P-phase. The N-phase extends into the region of substantially larger $ I $. 

The picture of  DOS for AFM subbands is shown in Fig.~\ref{fig:DOS_sc_and_bcc}b.
One can see  a strong van Hove singularity for the $\mathbf{Q}=\mathbf{Q}_{\rm H}$ point owing to nesting. 
Note that the singularity of the bare DOS does not survive here at finite $t'$.

\subsection{FCC lattice}
\begin{figure}[b]
\includegraphics[angle=-90,width=0.49\textwidth]{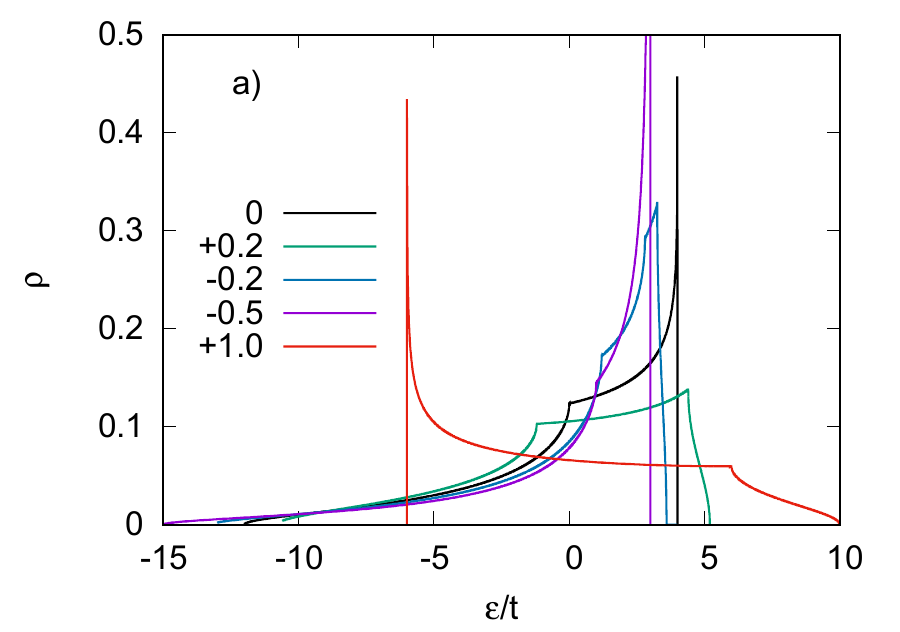}
\includegraphics[angle=-90,width=0.49\textwidth]{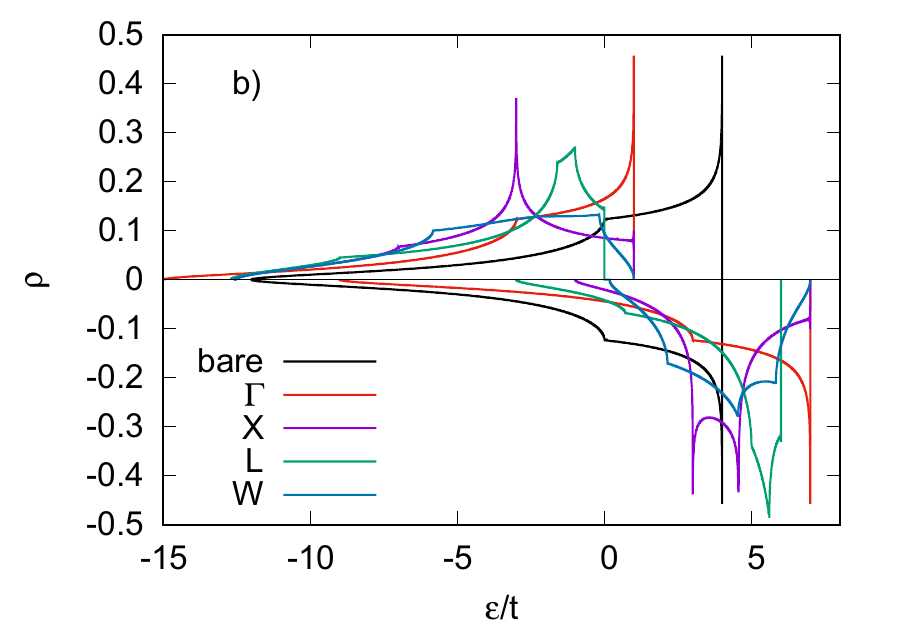}
\caption{a) The plot of DOS for fcc lattice bare spectrum at different $t'$, ratio $t'/t$ is shown in the legend. b) The same as in Fig.~\ref{fig:DOS_sq} at $I = 6t$ and $t' = 0$ for fcc lattice at $\mathbf{Q} = \mathbf{Q}_{\Gamma} = 0,\mathbf{Q}_{\rm X},\mathbf{Q}_{\rm L}, \mathbf{Q}_{\rm W}$.}
\label{fig:DOS_fcc}
\end{figure}

The plots of bare DOS for different $t'$ are shown in Fig.~\ref{fig:DOS_fcc}a, see also the analysis in Ref.~\onlinecite{2019:Igoshev_DOS_JETP}. The van Hove singularity lines and surfaces  produces  the following DOS singularities: logarithmic at $t' = 0$,  inverse fourth root at $t' = t$  and inverse square root at $t' =-t/2$, see~Ref.~\onlinecite{2019:Igoshev_DOS_JETP}. In Fig.~\ref{fig:DOS_fcc}b the DOS of AFM spectrum subbands at $t' = 0, I = 6t$ for various commensurate AFM orderings, $\mathbf{Q} = \mathbf{Q}_{\Gamma} = 0,\mathbf{Q}_{\rm X},\mathbf{Q}_{\rm L}, \mathbf{Q}_{\rm W}$ is shown, analogously to Figs.~\ref{fig:DOS_sq},\ref{fig:DOS_sc_and_bcc}.
 
\begin{figure} [ht]
\begin{minipage}[h]{0.59\linewidth}
\center{\includegraphics[angle=-90,width=0.99\linewidth]{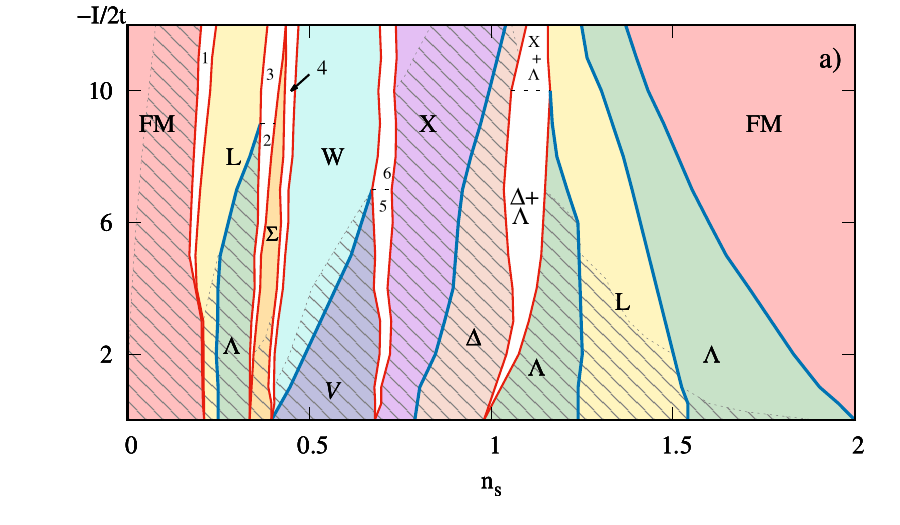}}
\end{minipage}
\begin{minipage}[h]{0.39\linewidth}
\center{\includegraphics[angle=-90,width=0.99\linewidth]{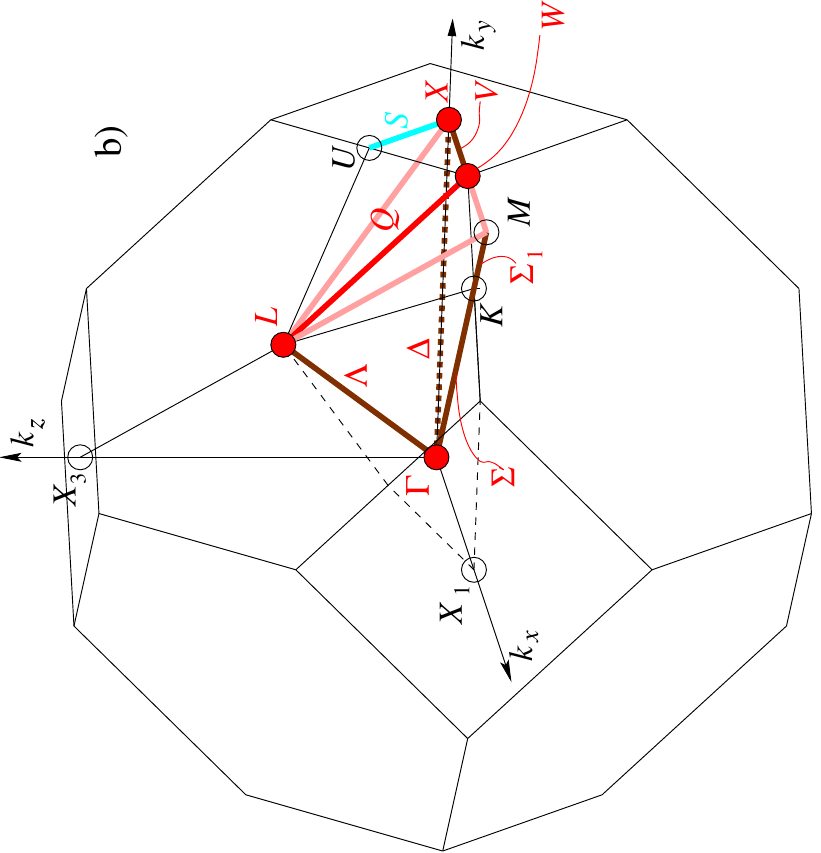}} 
\end{minipage}
\hfill
\begin{minipage}[h]{0.49\linewidth}
\center{\includegraphics[angle=-90,width=0.99\textwidth]{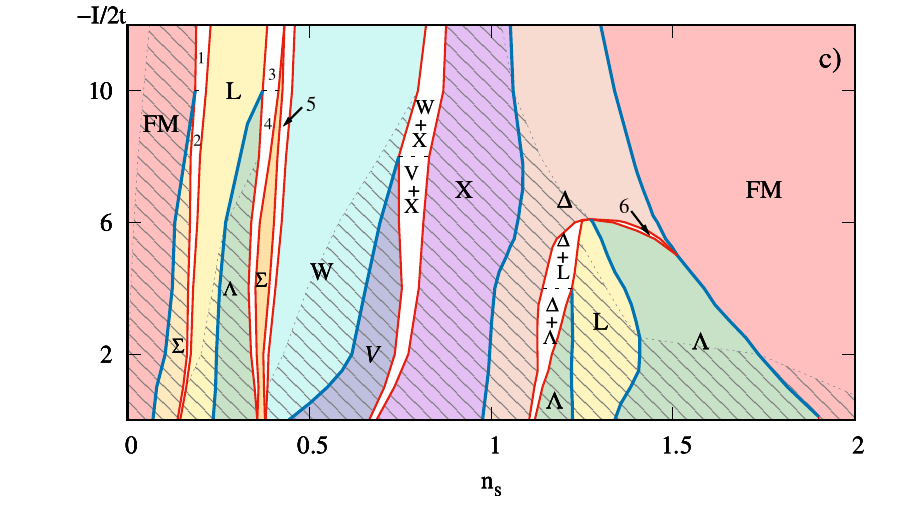} }
\end{minipage}
\hfill
\begin{minipage}[h]{0.49\linewidth}
\center{\includegraphics[angle=-90,width=0.99\textwidth]{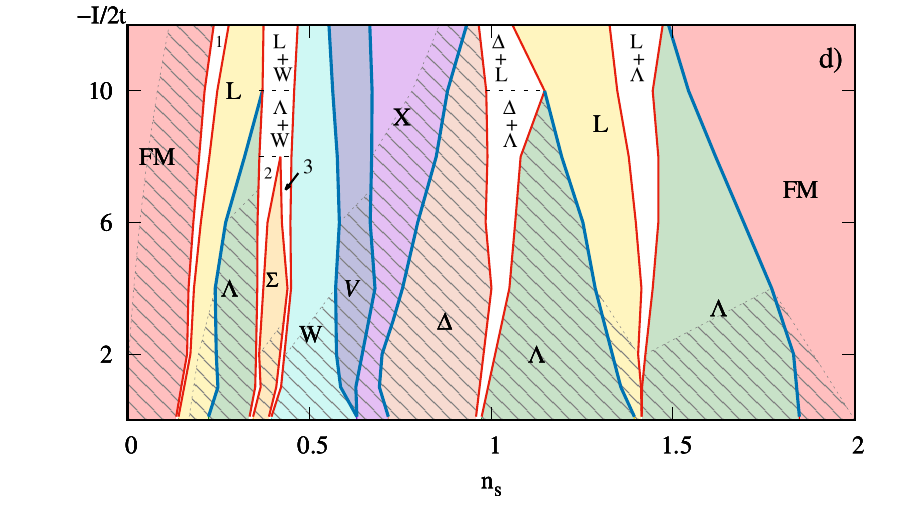}} 
\end{minipage}
\caption{The same as Fig. \ref{fig:sq} for the fcc lattice. a) $t' = 0$ with labels corresponding PS regions: 1 --- FM+L, 2 --- $\Lambda$+$\varSigma$, 3 --- L+$\varSigma$, 4 --- $\varSigma$+W, 5 --- \textit{V}+X, 6 --- W+X, b) the first Brillouin zone of the fcc lattice showing highly symmetric points and directions: $\Gamma(0,0,0)$, X$(2\pi,0,0)$, W$(2\pi,\pi,0)$, K$(3\pi/4,3\pi/4,0)$, L$(\pi,\pi,\pi)$, $\varDelta(q_x, 0, 0)$, $\varSigma(q_x, q_x, 0)$, $\varLambda(q_x, q_x, q_x)$, $V(2\pi,q_y,0)$, c)~$t'=-0.2t$ with labels corresponding PS regions: 1 --- FM+L, 2 --- $\varSigma$+L, 3 --- L+$\varSigma$, 4 --- $\Lambda$+$\varSigma$, 5 --- $\varSigma$+W, 6 --- $\Lambda$+$\varDelta$, d)~$t'=+0.2t$ with labels corresponding PS regions: 1 --- FM+L, 2 --- $\Lambda$+$\varSigma$, 3 --- $\varSigma$+W.}
\label{fig_fcc}
\end{figure}
Fig.~\ref{fig_fcc} shows the magnetic phase diagrams of the ground state for the fcc lattice. Since this is not bipartite, the cases of both signs of $t'$ should be considered. In all the  cases, a significant asymmetry of the region of ferromagnetic ordering is observed at $ n_s \ll 1 $ and $ 2 - n_s \ll 1 $: in the first case, the density of states, as usual for the three-dimensional situation, vanishes at the band bottom, and in the second a van Hove singularity appears at the band top at $ t'=0 $. At $ t'\ne 0 $, the shape of DOS near the band top is as follows: there is a sharp peak remaining from the van Hove singularity at $ t'=0 $, which abruptly vanishes at the band top, see~Fig.~\ref{fig:DOS_fcc}a. (Such a sharp drop can lead to interesting effects in electronic properties~\cite{2019:Igoshev_DOS_JETP,2019:Igoshev_DOS_FMM}).  Thus the rest of the peak in DOS generally leads to a wide region of saturated ferromagnetic ordering at $ n_s $ near 2.
With  further deviation of $ n_s $ from 2, the ferromagnetic order is replaced by a ``diagonal'' $ \varLambda $-order (the $ \Gamma $ L direction in the~Brillouin zone). 
It is interesting that in the case $ n_s \gtrsim 1 $ the ferromagnetic order,  collinear antiferromagnetic order L (order of the second kind, type I~\cite{book:Goodenough}) and the close  $ \varLambda $ AFM orders  occupy almost half of the entire phase diagram (only for $ t '= -0.2t $, there is also a region of $ \varDelta $-order, which is also close to ferromagnetic).  
The case $ n_s = 1 $ is not distinguished, since there is no nesting of the Fermi surface in this case. Therefore, depending on the value of $ t'$, completely different cases can be realized. In general, we can conclude that an increase of $ t'$ leads to a shift of the phase regions for the phases X and $ \varDelta $ to the right, displacing the $ \varLambda $-phase.
At $ t'=0 $, an unsaturated $ \Delta $ spiral order is formed, and at $ t'= -0.2t $ unsaturated $ \varDelta $-order (for small $ I $) and X-order (for moderate and large $ I $) which is collinear AFM order of the first kind according to  \cite{book:Goodenough}; at $ t '= + 0.2t $, the phase separation occurs into unsaturated $ \varDelta $ ($ n_s <1 $) and $ \varLambda $ ($ n_s> 1 $) phases.

For $ n_s \sim 0.5 $, the formation of a spiral magnetic order with the wave vector located on the square edge of the Brillouin zone is characteristic~(X---$V$---W).

\subsection{Discussion}

Within the framework of the itinerant electron  Hubbard model, which does not include localized moments into consideration, the phase diagrams of the ground state were determined in a similar approximation,  the magnetic order types and magnetic phase separation~\cite{2010:Igoshev,2015:Igoshev} being investigated. Note that the substitution of $ I = U_dm_s $, where $U_d$ is on-site Coulomb interaction parameter, into the equations for the parameters of the phases (\ref{eq:ns}), (\ref{eq:ms}) and the grand potential (\ref{eq:omega}) 
reproduces the corresponding equations of the Hubbard model, provided that the saturated state is considered ($ n_s = 2m_s $). In  particular, for the square lattice the sequence of magnetic phase transitions with deviation from half-filling ($n_s=1$), M (N\'eel antiferromagnetism)$\rightarrow Z\rightarrow$X$\stackrel{\rm PS}{\rightarrow}\varDelta\rightarrow\Gamma$~(ferromagnetism), is the same in the $s-d$ exchange and Hubbard model, see Fig.~1 of Ref.~\onlinecite{2018:Igoshev_FMM} and Fig.~\ref{fig:sq}.

However, a direct comparison of $ \Omega $ for different phases specified by the wave vector for the $ s $-$ d (f) $ model is not equivalent to such a comparison for the Hubbard model in the case when a first-order transition between these phases is considered (for the parameter values inside the region of phase separation), since  the first-order phase transition in the Hubbard model between saturated phases with~$ m = m_1 $ and~$ m = m_2 $ corresponds to different $ I_1 = U_dm_1 $ and $ I_2 = U_dm_2 $ in the for $ s $-$ d (f) $ model. At the same time, consideration of a second-order transition between saturated AFM and FM phases will give the same result for both models.

The presence of localized electrons contributes to the stabilization of the FM order at a low concentration of current carriers even at small values of $ I $, and for the case of a large number of carriers the results are qualitatively similar: the states found and the types of phase transitions between them are the same.

According to the Schrieffer-Wolff transformation~\cite{SW}, the Hamiltonian of the Anderson model in the limit of small hybridization of $ s $- and $ d (f) $-states,  $ | V | \ll | \epsilon_d- \epsilon _{\rm F} |, | U_d + \epsilon_d- \epsilon _{\rm F} | $ 
(where $ \epsilon_d $ is the energy level of the $ d $-state, $\epsilon_{\rm F}$ is the Fermi level of $s$-electron subsystem, $V$ is $s$-$d(f)$ hybridization interaction parameter),
reduces to $ s $-$ d (f) $ model with integer filling of $d(f)$-level and the exchange integral 
\begin{equation}
	I = V^2\left(\frac1{\epsilon_d-\epsilon_{\rm F}} - \frac1{U_d+\epsilon_d-\epsilon_{\rm F}}\right).
\end{equation}
The calculation results ~\cite{2016:Igoshev} for the parameters $|V/U_d|$~($\sim 10^{-2}$) and $|V/\epsilon_d|$($\sim 10^{-1}$) for a square lattice in  the nearest-neighbor approximation completely agrees with the present calculations. However, when the conditions for the applicability of the Schrieffer-Wolf transformation are violated, the phase diagram in the intermediate valence regime changes significantly, especially for a situation far from half-filling: a narrow ferromagnetic phase appears only for one sign of the carriers; for the opposite one, there is a wide region of phase separation with the participation of antiferromagnetic phases. It is interesting that the ferromagnetic phase appears at sufficiently large values of the hybridization parameter $ V $ upon intermediate doping.

Recently, the Kondo screening of the magnetic moments   in  the $ s $-$ d (f) $ model has been investigated  for a square lattice in the nearest-neighbor approximation within the mean-field approximation  introducing Abrikosov fermion representation for localized spins~\cite{Costa}. However, the phase separation, which necessarily arises in first-order transitions, was not taken into account. 
For the case of not too large $ I \lesssim 2t $, these results are in good agreement with those obtained by us, if we disregard the phase separation. However, at large $I$, the Kondo screening of the local moment begins to play an important role.
It is interesting that the values of the $ I $  at which the Kondo screening becomes significant just correspond  to the transition from  incommensurate to  commensurate magnetic order.

The formation of magnetic order for the $s-d(f)$~model was considered in the limit of strong exchange interaction~$I$, which leads to the realization of the double exchange mechanism~\cite{Izumov:DE} and ferromagnetic ordering~\cite{1992:Sigrist}. 

The opposite case of moderate exchange interaction in the nearest-neighbor approximation was considered for a square lattice taking into account additional competing direct-exchange antiferromagnetic interaction between localized spins~\cite{1995:Hamada}. The result of this competition in the case of a small number of carriers is a canted magnetic order, which should be in fact replaced by phase separation, as demonstrate our calculations. In the case of a larger number of carriers, a type of spiral or antiferromagnetic order is realized.

\section{Conclusion}
The results obtained for various lattices demonstrate a great variety of magnetic structures and phase transitions, which is significantly more rich than for the Heisenberg magnets. These results are quite general and make it possible to interpret magnetic properties and describe the sequence of magnetic phase transitions for  compounds of  $d$-metals, in particular based on iron, manganese and chromium~\cite{2016:Review_MnP} (see also Introduction).
 
 There exist examples of systems with magnetic phase transitions of the type ferromagnet--incommensurate order: MnP with orthorhombic (distorted hexagonal) lattice, the wave vector of the incommensurate phase being directed along the $x$ axis~\cite{1966:MnP,1974:Helimagnetism,2016:MnP}; YbAgGe (the lattice is hexagonal noncentrosymmetric, the wave vectors are $(1/3,0,1/3)\rightarrow(0,0,0.324)$~\cite{YbAgGe,2004:Fak,2006:Fak}; CePtAl (orthorhombic crystal lattice, magnetic order is a superposition of ferromagnetic and incommensurate order  with $\mathbf{Q}=(0,0.463(8),0)$, and of commensurate AFM order with $\mathbf{Q}=(0,0.5,0)$)~\cite{1995:CePtAl}.
 
We have found commensurate types of magnetic ordering (ferromagnetic and antiferromagnetic), transitions between  with a change in the number of carriers $ n_s $, which are realized through regions of phase separation and  spiral phases. Spiral phases at small $ I $ occupy most of the phase diagram, and, as $ I $ increases, they are displaced by commensurate phases. Near the top and bottom of the band, a ferromagnetic order is found for each diagram.

The type of magnetic ordering and the nature of phase transitions strongly depend on the geometric properties of the crystal lattice. For bipartite lattices with the nesting property (square, simple cubic, and bcc), a checkerboard order for half filling and a commensurate order of another type for a significant deviation  from half filling dominate. If the deviation of the carrier concentration from the half-filling is not too large, phase separation is realized into staggered and incommensurate (for small $ I $) or non-staggered orders for larger $I$. Thus, a commensurate antiferromagnetic order is considerably protected. The antiferromagnetic order has the saturation property. For a small number of carriers, ferromagnetic order is realized, the region of the ferromagnetic phase being narrow far from the conditions of applicability of the double exchange model ($W\ll IS$).

For the fcc lattice, the situation is  different: all magnetic phase transitions are of either  the second order or a weak first order (a narrow stripe of phase separation). At filling  near the top of the band, saturated ferromagnetism occupies a wide region, which is again due to the strong logarithmic van Hove singularity. The antiferromagnetic order has a significantly different spatial configuration near half-filling, depending on the carrier concentration. Thus in this case the magnetic order is strongly degenerate, and it can be expected that the type of antiferromagnetic order  will strongly depend on doping, pressure, and spatial deformation. This is not the case for $n_s\gtrsim 1.5$, when the ferromagnetic and long-wavelength diagonal incommensurate orders have significantly lower energies than the regular antiferromagnetic order. 

The saturation of the magnetic state depends essentially  on the dimensionality of the system and its bipartite property. For a square lattice, the antiferromagnetic order is almost always saturated due to the presence of a strongly van Hove singularity, while for simple cubic and bcc lattices, an unsaturated antiferromagnetic order is formed only at small $ I $. For a non-bipartite fcc lattice, the antiferromagnetic order is generally unsaturated (except for the case where the Fermi level is close to the band top or very large $I$). 

Not all types of collinear antiferromagnetism \cite{book:Goodenough}, e.g. E-type AFM ordering for HoMnO$_3$ and YMnO$_3$, as well as nonsymmorphic symmetric ordering, which are often encountered in antiferromagnetic materials, can be studied within the framework of the approach used here. However, the direct introduction of magnetic sublattices into the formalism of calculations being combined with the spiral form of magnetic order can, in the future, completely solve the problem of describing all these orders.

The research was carried out within the Ministry State Assignment  of Russia (theme ``Quantum'' No. AAAA-A18-118020190095-4).


\end{document}